\documentclass[11pt,a4paper]{article}
\pdfoutput=1
\usepackage{jheppub}
\usepackage{tikz}
\usetikzlibrary{intersections, calc, arrows.meta, bending}
\usepackage{subcaption}
\usepackage{simpler-wick}
\usepackage{amsmath,amssymb,mathtools}

\DeclareMathOperator{\Tr}{Tr}

\newcommand{\ri}{\mathrm{i}}
\renewcommand{\th}{\theta}

\newcommand{\cob}{\delta}

\newcommand{\vep}{\varepsilon}

\newcommand{\hf}{\frac{1}{2}}

\newcommand{\til}[1]{\widetilde{#1}}

\newcommand{\Si}{\Sigma}

\newcommand{\del}{\partial}

\newcommand{\lap}{\Delta}
\newcommand{\bra}{\langle}
\newcommand{\ket}{\rangle}
\newcommand{\la}{\lambda}

\newcommand{\h}[1]{\widehat{#1}}
\newcommand{\bt}{\beta}

\newcommand{\rt}[1]{\sqrt{#1}}
\newcommand{\cO}{\mathcal{O}}

\newcommand{\cD}{\mathcal{D}}

\newcommand{\cA}{\mathcal{A}}

\def\nn{\nonumber}

\begin{document}

\begin{flushright}
{RUP-23-5}
\end{flushright}

\title{Correlators of double scaled SYK at one-loop}

\author[a]{Kazumi Okuyama}
\author[b]{and Kenta Suzuki}

\affiliation[a]{Department of Physics, 
Shinshu University, 3-1-1 Asahi, Matsumoto 390-8621, Japan}
\affiliation[b]{Department of Physics, Rikkyo University, Toshima, Tokyo 171-8501, Japan}

\emailAdd{kazumi@azusa.shinshu-u.ac.jp, kenta.suzuki@rikkyo.ac.jp}

\abstract{
In this paper, we study one-loop contributions in the double-scaling limit of the SYK model from the chord diagrams and Liouville type effective action.
We compute and clarify the meaning of each component consisting of the one-loop corrections for the two- and time-ordered four-point functions of light operators.
We also reproduce the exact expression of the out-of-time-ordered four-point function at arbitrary temperatures within the one-loop level,
which were previously computed from different methods.

}

\maketitle

\section{Introduction}
\label{sec:introduction}
The large $N$ dynamics of the Sachdev-Ye-Kitaev (SYK) model \cite{Sachdev1993,Kitaev1,Kitaev2,Polchinski:2016xgd,Maldacena:2016hyu} represents a particularly useful laboratory to understand the origins of the AdS/CFT correspondence.
In the low temperature limit, the crucial role is played by the Schwarzian mode which is responsible for the breaking of the emergent IR reparametrization symmetry.
This Schwarzian mode triggers the maximal chaos behaviour \cite{Maldacena:2015waa} in the low temperature.
This mode also builds a connection with the Jackiw-Teitelboim (JT) gravity through the near-AdS$_2$/near-CFT$_1$ correspondence \cite{Almheiri:2014cka, Jensen:2016pah, Maldacena:2016upp, Engelsoy:2016xyb}, where breaking of the diffeomorphism in AdS$_2$ also leads to an emergence of the Schwarzian mode.

Although the SYK model led us to a great deal of understanding of the AdS/CFT correspondence, most of the previous works on this model were limited to low temperature (near conformal) limit, and one might wish to go beyond this limit.
The large $p$ limit \cite{Maldacena:2016hyu,Cotler:2016fpe,Tarnopolsky:2018env,Das:2020kmt} and the double-scaling limit \cite{Cotler:2016fpe,Berkooz:2018qkz,Berkooz:2018jqr,Lin:2022rbf,Okuyama:2022szh,Goel:2023svz} of the SYK model give one possibility in this direction.
Previous studies on the large $p$ SYK model showed a transition from the maximal chaos bound to non-maximal chaos behaviour \cite{Maldacena:2016hyu},
effective Liouville type action \cite{Cotler:2016fpe} and corrections on top of that action \cite{Das:2020kmt}.
The large $p$ limit also allows us to investigate modifications away from the conformal two-point function \cite{Tarnopolsky:2018env}
as well as out-of-time-order four-point functions \cite{Streicher:2019wek,Choi:2019bmd,Gu:2021xaj}.
The double-scaling limit of the SYK model studied in \cite{Berkooz:2018qkz,Berkooz:2018jqr,Lin:2022rbf,Okuyama:2022szh,Berkooz:2020uly,Berkooz:2020xne}
uses a particular method called chord diagrams, which we review in section~\ref{sec:DSSYK}, and leads to an exact results for the partition function and two- and four-point functions.
However, the bulk gravitational dual of the DSSYK model is not well understood, if it exists, and it is desirable to investigate further in this direction. 
Some recent works toward this direction includes \cite{Goel:2023svz,Berkooz:2022mfk,Mukhametzhanov:2023tcg}. 
In order to contribute for this investigation, we study one-loop corrections in the DSSYK model in this paper.

The rest of the paper is organized as follows.
In section~\ref{sec:DSSYK}, we give a brief review of the DSSYK model with two formalisms.
One is based on the chord diagrams and the other is formulated by a Liouville type effective action.
In section~\ref{sec:saddle}, we study one-loop corrections for the two- and uncrossed four-point functions in the DSSYK model,
starting from the results obtained by the chord diagram method.
Some of these one-loop corrections were partially studied in \cite{Goel:2023svz} as well, but we explain the meaning of each components consisting of the one-loop contributions.
In section~\ref{sec:Liouville}, we study one-loop corrections for the two- and uncrossed four-point functions starting from the Liouville type effective action,
and reproduce the results found in section~\ref{sec:saddle}.
In section~\ref{sec:OTOC}, we continue our study of the Liouville theory for the out-of-time-ordered four-point functions and we reproduce the results obtained in \cite{Streicher:2019wek,Choi:2019bmd,Gu:2021xaj} by different methods.
In section~\ref{sec:low}, we compare our results obtained in the previous section with the known results from the low temperature Schwarzian theory.
Finally, we conclude in section \ref{sec:conclusion} with some discussion of future directions.
In appendix~\ref{app:alt}, we present an alternative derivation of the one-loop determinant based on a change of variables which diagonalizes the Hessian.
In appendix~\ref{app:sum}, we summarize useful summation formulae used in the main text.
In appendix~\ref{app:factorization}, we discuss zero temperature factorization of the four-point function into a pair of two-point functions in the DSSYK model.

\section{Review of double scaled SYK}
\label{sec:DSSYK}
In this section, we give a brief review of the double-scaled SYK model.
The Sachdev-Ye-Kitaev model \cite{Sachdev1993, Kitaev1, Kitaev2} is a quantum mechanical many body system with all-to-all $p$-body interactions on fermionic $N$ sites ($N \gg 1$), represented by the Hamiltonian
	\begin{align}
		H \, = \, \ri^{\frac{p}{2}} \sum_{1 \le i_1 < \cdots < i_p \le N} J_{i_1 i_2 \cdots i_p} \, \psi_{i_1} \, \psi_{i_2} \, \cdots \psi_{i_p} \, ,
	\end{align}
where $\psi_i$ are Majorana fermions, which satisfy $\{ \psi_i, \psi_j \} = 2\delta_{ij}$.
The coupling constant $J_{i_1 i_2 \cdots i_p}$ is random with a Gaussian distribution
	\begin{align}
		\qquad \big\langle J_{i_1 i_2 \cdots i_p} \big\rangle_J \, = \, 0 \, , \qquad
        \big\langle J_{i_1 i_2 \cdots i_p}^2 \big\rangle_J \, = \, \binom{N}{p}^{-1} \quad (\textrm{no\ sum\ for\ indices}) \, .
	\label{eq:J}
    \end{align}
The double-scaled SYK (DSSYK) model is defined by taking the double scaling limit
	\begin{align}
		\qquad \qquad N , \, p \, \to \, \infty \quad \textrm{with} \quad \lambda \, := \, \frac{2p^2}{N} \quad \textrm{fixed} \, .
	\label{eq:double_scaling}
    \end{align}
There are two ways to study the DSSYK.
The one is by the chord diagrams \cite{Berkooz:2018qkz,Berkooz:2018jqr} which leads to complicated but exact results.
The other is by the $G\Sigma$ formalism \cite{Cotler:2016fpe} which leads to a Liouville type action well suited for small $\lambda$ perturbations.
In this paper, we study both formalisms.

\begin{figure}
    \begin{center}
    \includegraphics[width=42mm]{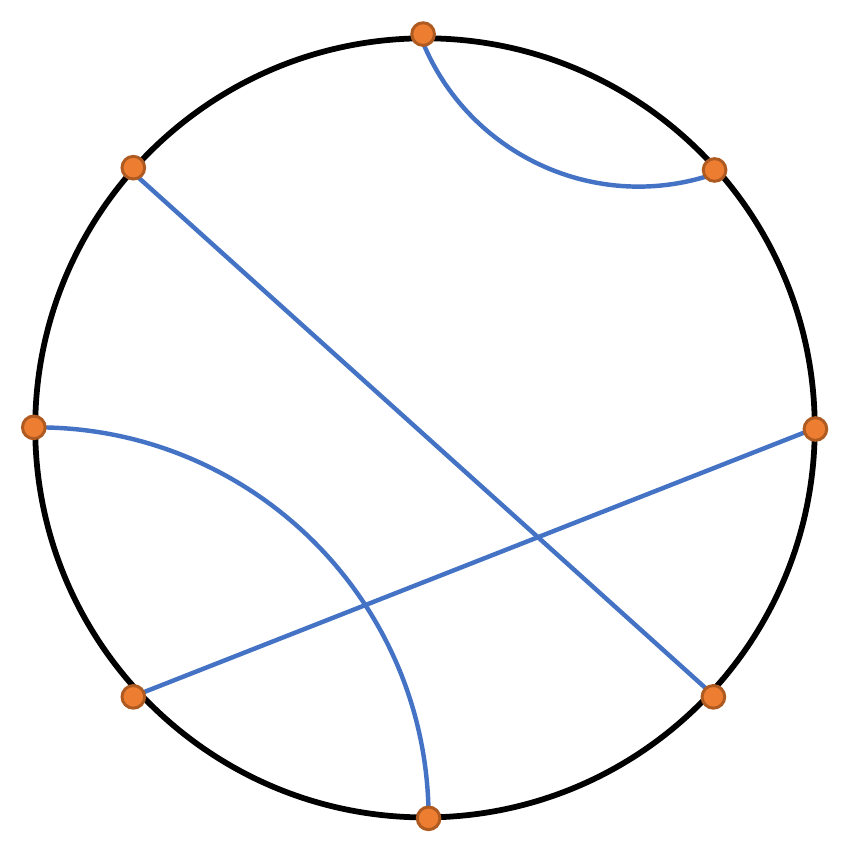}
    \end{center}
    \caption{An example of a chord diagram for $k=8$. The black circle represents a trace over the fermions.
    Each orange dot represents an insertion of a Hamiltonian and each blue line denotes a contraction of capital index $I$. }
    \label{fig:chord}
\end{figure}

The chord diagram method consider, for example, the disorder averaged partition function (but the same method also works for correlation functions as well)
	\begin{align}
		\big\langle Z \big\rangle_J \, := \, \big\langle \Tr e^{- \beta H} \big\rangle_J \, ,
	\end{align}
and expands $e^{-\beta H}$ to rewrite in terms of summation over the moments $m_k$
	\begin{align}
		\big\langle Z \big\rangle_J \, = \, \sum_{k=0}^{\infty} \frac{(-\beta)^k}{k!} \, m_k \, , \qquad m_k \, := \, \big\langle \Tr \big( H^k \big) \big\rangle_J \, .
	\end{align}
The evaluation of the moments is reduced to a product between a trace over fermions and disorder average over the coupling constants as
	\begin{align}
		m_k \, = \, \ri^{\frac{kp}{2}} \, \big\langle J_{I_1} \cdots J_{I_k} \big\rangle_J \Tr(\psi_{I_1} \cdot \psi_{I_k}) \, ,
	\end{align}
where the capital index $I$ represents a  set of $p$ indices $i_1 \cdots i_p$ and $\psi_I = \psi_{i_1} \cdots \psi_{i_p}$.
Evaluation of the disorder average over the coupling constants by (\ref{eq:J}) leads to contraction of the indices.
Then the trace over the fermions is represented by the chord diagrams (for example see Figure~\ref{fig:chord} for $k=8$ chord diagram.)
After resummining over the moments, the disorder averaged partition function is found as \cite{Berkooz:2018qkz,Berkooz:2018jqr}
	\begin{align}
		\big\langle Z \big\rangle_J \, = \, \int_0^\pi \frac{d\theta}{2\pi} \, \mu(\theta) e^{-\beta E(\theta)} \, ,
	\end{align}
where 
	\begin{align}
		E(\theta) \, = \, - \frac{2\cos\theta}{\sqrt{1-q}} \, , \qquad q \, := \, e^{-\lambda} \, ,
  \label{eq:E-theta}
	\end{align}
and 
	\begin{align}
		\mu(\theta) \, = \, (q;q)_\infty \, (e^{2\ri \theta}; q)_\infty \, (e^{-2\ri \theta}; q)_\infty \, .
	\end{align}
Here the $q$-Pochhammer symbol is defined by
	\begin{align}
		(a;q)_n \, = \, \prod_{k=0}^{n-1} (1-a q^k) \, .
	\end{align}
Similarly, using the chord diagram method, the two-point function of operators with dimension $\Delta$
    \begin{align}
         \til{G}_2(\beta_1, \beta_2) \, = \, \int_0^\pi\prod_{k=1,2}\frac{d\th_k}{2\pi}\mu(\th_k) e^{-\bt_kE(\th_k)}\frac{(e^{-2\lap};q)_\infty}{(e^{-\lap+\ri(\pm\th_1\pm\th_2)};q)_\infty} \, ,
    \label{eq:G2_intro}
    \end{align}
the uncrossed four-point function of pairs of operators with dimension $\Delta_1$ and $\Delta_2$
    \begin{equation}
        \til{G}_4(\beta_1, \beta_2, \beta_3, \beta_4) \, = \, \int\prod_{i=1}^3 \left( \frac{d\th_i}{2\pi}\mu(\th_i)e^{-\bt_i E(\th_i)} \right)
        \frac{(e^{-2\lap_1};q)_\infty}{(e^{-\lap_1+\ri(\pm\th_1\pm\th_3)};q)_\infty}
        \frac{(e^{-2\lap_2};q)_\infty}{(e^{-\lap_2+\ri(\pm\th_2\pm\th_3)};q)_\infty} \, ,
    \label{eq:G4_intro}
    \end{equation}
are obtained.
By $(e^{-\lap+\ri(\pm\th_1\pm\th_2)};q)_\infty$, we mean to take a product of all four possible combination of the signs.
The crossed four-point function is also obtained by the chord diagram method in \cite{Berkooz:2018jqr},
but we do not write the result explicitly here since we don't use it.
We study small $\lambda$ expansions of these results obtained from chord diagrams in section~\ref{sec:saddle}.

The other $G\Sigma$ formalism, we introduce bi-local fields $G(\tau_1, \tau_2)$ and $\Sigma(\tau_1, \tau_2)$
as in the usual Hubbard–Stratonovich transformation to integrate out the original fermions.
Then, the disorder averaged partition function is written as
	\begin{align}
		\big\langle Z \big\rangle_J \, = \, \int \cD G \cD\Sigma \, e^{-S} \, ,
	\end{align}
where 
	\begin{align}
		S \, = \, - \, \frac{N}{2} \log \det(\partial_\tau - \Sigma) \, + \, \frac{N}{2} \int d\tau_1 d\tau_2 \left[ \Sigma G - \frac{1}{2p^2} (2G)^p \right] \, .
	\end{align}
In order to take the double scaling limit (\ref{eq:double_scaling}), we set
	\begin{align}
		G(\tau_1, \tau_2) \, = \, \frac{\textrm{sgn}(\tau_{12})}{2}\left( 1 + \frac{g(\tau_1, \tau_2)}{p} \right) \, , \qquad 
        \Sigma(\tau_1, \tau_2) \, = \, \frac{\sigma(\tau_1, \tau_2)}{p} \, , 
	\end{align}
where $\tau_{12}:=\tau_1 - \tau_2$.
Substituting this expression into the action, one can integrate out the $\sigma$ field in the leading order of large $p$.
Hence, in the double scaling limit, we find the action given by
	\begin{align}
		S \, = \, \frac{1}{2\lambda} \int d\tau_1 d\tau_2 \left[ \, \frac{1}{4} \, \partial_1 g(\tau_1, \tau_2) \partial_2 g(\tau_1, \tau_2) \, - \, e^{g(\tau_1, \tau_2)} \right] \, .
	\end{align}
We study this small $\lambda$ Liouville theory in section~\ref{sec:Liouville} and \ref{sec:OTOC}.

\section{Saddle point computation of correlators}
\label{sec:saddle}
The small $\la$ regime of DSSYK corresponds to the semi-classical
bulk gravitational theory. In \cite{Goel:2023svz}
the small $\la$ expansion of the matter correlators of DSSYK
was computed up to the one-loop order.
In this section, we will generalize the analysis in \cite{Goel:2023svz}
and compute the one-loop correction to the uncrossed four-point function.

In order to take a well-defined small $\la$ limit, 
we have to rescale the inverse temperature as $\bt\to\bt/\rt{\la}$. 
Equivalently, we can rescale $E(\th)$ in \eqref{eq:E-theta} as
\begin{equation}
\begin{aligned}
E(\th)=-\frac{2\cos\th}{\rt{\la(1-q)}},
\end{aligned} 
\end{equation}
with $\bt$ intact.
We will use this convention throughout the rest of this section.

\subsection{Partition function}
Let us first consider the small $\la$ expansion of the partition function.
As discussed in \cite{Goel:2023svz}, this expansion is obtained from
the saddle point approximation of the $\th$-integral.
To do that, we need the small $\la$ expansion of the $q$-Pochhammer symbol
\begin{equation}
\begin{aligned}
 (a;q)_\infty&=\exp\left[-\sum_{n=1}^\infty\frac{a^n}{n(1-q^n)}\right]
=\exp\left[-\sum_{g=0}^\infty \frac{\la^{2g-1}B_{2g}}{(2g)!}
\text{Li}_{2-2g}(a)+\hf\log(1-a)\right],
\end{aligned} 
\end{equation}
where $B_{2g}$ denotes the Bernoulli number and 
$\text{Li}_n(z)$ is the polylogarithm.
The small $\la$ expansion of $(q;q)_\infty$
is also obtained by using its relation to the Dedekind $\eta$-function
\begin{equation}
\begin{aligned}
(q;q)_\infty=q^{-\frac{1}{24}} \eta(q)\approx\rt{\frac{2\pi}{\la}}e^{\frac{\la}{24}-\frac{\pi^2}{6\la}},
\end{aligned} 
\end{equation} 
where in the last step we used the S-transformation of the $\eta$-function.
Using the relation
\begin{equation}
\text{Li}_{2-2g}(e^{2\ri\th})+\text{Li}_{2-2g}(e^{-2\ri\th})=
\left\{
\begin{aligned}
&2\left(\th-\frac{\pi}{2}\right)^2-\frac{\pi^2}{6},\qquad&(g=0),\\
&-1,\qquad&(g=1),\\
 &0,\qquad &(g\geq2),
\end{aligned}\right.
\end{equation}
the measure factor $\mu(\th)$ is expanded as
\begin{equation}
\begin{aligned}
 \mu(\th)&=(q;q)_\infty (e^{\pm2\ri\th};q)_\infty
\approx\rt{\frac{2\pi}{\la}}\exp\left[\frac{\la}{8}-\frac{2}{\la}\left(\th-\frac{\pi}{2}\right)^2+\log(2\sin\th)\right].
\end{aligned} 
\label{eq:mu-expand}
\end{equation}
Note that there is no
corrections to $\mu(\th)$ higher than $\cO(\la^2)$.\footnote{
The same conclusion can be obtained by using the expression of
$\mu(\th)$ in terms of the Jacobi theta-function
$\vartheta_1(z,\tau)$
\begin{equation}
\begin{aligned}
 \mu(\th)=2 q^{-\frac{1}{8}}\sin\th\,\vartheta_1\left(2\th,\frac{\ri\la}{2\pi}\right).
\end{aligned} 
\end{equation}
}
Then the partition function up to $\cO(\la)$ is written as
\begin{equation}
\begin{aligned}
 Z(\bt)&=\int_0^\pi\frac{d\th}{2\pi}\mu(\th)e^{-\bt E(\th)}
\approx\rt{\frac{2\pi}{\la}}\int_0^\pi\frac{d\th}{2\pi}e^{-\frac{1}{\la}F+h},
\end{aligned} 
\label{eq:Z-int}
\end{equation}
where 
\begin{equation}
\begin{aligned}
 F&=2\left(\th-\frac{\pi}{2}\right)^2-2\bt\cos\th,\\
h&=\log(2\sin\th)+\hf \bt\cos\th.
\end{aligned} 
\end{equation}
In the small $\la$ limit, the $\th$-integral in \eqref{eq:Z-int}
is evaluated by the saddle point approximation.
The saddle point $\th=\th_*$ is determined from the saddle
point equation $\del_\th F=0$ as
\begin{equation}
\begin{aligned}
 \th_*=\frac{\pi}{2}-u,
\end{aligned} 
\label{eq:th*}
\end{equation}
where $u$ is related to $\bt$ as
\begin{equation}
\begin{aligned}
 \bt=\frac{2u}{\cos u}.
\end{aligned}
\label{eq:saddle-u} 
\end{equation}
The saddle point value of $F$ is
\begin{equation}
\begin{aligned}
 F_*=F(\th_*)=2(u^2-2u\tan u).
\end{aligned} 
\label{eq:f0}
\end{equation}
Note that our $u$ and $v$ in \cite{Maldacena:2016hyu} are related by
\begin{equation}
\begin{aligned}
u=\frac{\pi v}{2}.
\end{aligned} 
\label{eq:u-v}
\end{equation}  

One can systematically improve the approximation by expanding the 
integral around the saddle point
\begin{equation}
\begin{aligned}
 \th=\th_*+\rt{\la}\vep.
\end{aligned} 
\end{equation}
Expanding $F$ up to the quadratic order in $\vep$, we find
\begin{equation}
\begin{aligned}
 \frac{1}{\la}(F-F_*)=2(1+u\tan u)\vep^2+\cO(\vep^3).
\end{aligned} 
\label{eq:vep2}
\end{equation}
Then, by performing the Gaussian integral over $\vep$, we find 
the one-loop correction to the partition function \cite{Goel:2023svz}
\begin{equation}
\begin{aligned}
 Z(\bt)
\approx \frac{\cos u}{\rt{1+u\tan u}}\exp\left[
-\frac{2}{\la}\Bigl(u^2-2u\tan u\Bigr)+u\tan u\right].
\end{aligned} 
\label{eq:Z1loop}
\end{equation}

\subsection{Two-point function}
The two-point function is given by \cite{Berkooz:2018jqr}
\begin{equation}
\begin{aligned}
 \til{G}_2
&=\int_0^\pi\prod_{k=1,2}\frac{d\th_k}{2\pi}\mu(\th_k)
e^{-\bt_kE(\th_k)}\frac{(e^{-2\lap};q)_\infty}{(e^{-\lap+\ri(\pm\th_1\pm\th_2)};q)_\infty}.
\end{aligned} 
\label{eq:tildeG2}
\end{equation}
We have put tilde to indicate that the two-point function 
is not normalized by the partition function. We define the normalized 
$n$-point function by
\begin{equation}
\begin{aligned}
 G_n=\frac{\til{G}_n}{Z}.
\end{aligned} 
\label{eq:normalizedG}
\end{equation}
$\bt_{1,2}$ in \eqref{eq:tildeG2} are given by
\begin{equation}
\begin{aligned}
 \bt_1=\tau_{12},\quad \bt_2=\bt-\tau_{12},
\end{aligned} 
\label{eq:bt-tau}
\end{equation}
with $\tau_{ij}=\tau_i-\tau_j$.
In the small $\la$ limit, \eqref{eq:tildeG2}
 is written as
\begin{equation}
\begin{aligned}
 \til{G}_2=\frac{2\pi}{\la}\int\prod_{k=1,2}\frac{d\th_k}{2\pi}
e^{-\la^{-1}F+h+\cO(\la)},
\end{aligned} 
\end{equation}
where $F$ and $h$ are given by
\begin{equation}
\begin{aligned}
F&=\sum_{k=1,2}\left[2\left(\th_k-\frac{\pi}{2}\right)^2-2\bt_k\cos\th_k\right]
+\text{Li}_2(e^{-2\lap})-\text{Li}_2(e^{-\lap+\ri(\pm\th_1\pm\th_2)}),\\
 h&=\sum_{k=1,2}\left[\hf\bt_k\cos\th_k+\log(2\sin\th_k)\right]
+\hf\log(1-e^{-2\lap})-\hf\log(1-e^{-\lap+\ri(\pm\th_1\pm\th_2)}).
\end{aligned} 
\label{eq:Fh-2pt}
\end{equation}
Note that the last term of $F$ and $h$ are written as
\begin{equation}
\begin{aligned}
-\text{Li}_2(e^{-\lap+\ri(\pm\th_1\pm\th_2)}) &=
-4\sum_{n=1}^\infty\frac{e^{-\lap n}}{n^2}\cos(n\th_1)\cos(n\th_2),\\
-\hf\log(1-e^{-\lap+\ri(\pm\th_1\pm\th_2)})&=
-\hf\log\Bigl[1-2e^{-\lap}\cos(\th_1+\th_2)+e^{-2\lap}\Bigr]\\
&\quad-\hf\log\Bigl[1-2e^{-\lap}\cos(\th_1-\th_2)+e^{-2\lap}\Bigr].
\end{aligned} 
\end{equation}

We would like to evaluate $\til{G}_2$ by the saddle point approximation.
The saddle point equation reads
\begin{equation}
\begin{aligned}
 \hf\frac{\del F}{\del\th_1}&=2\th_1-\pi+\bt_1\sin\th_1
+2\sum_{n=1}^\infty \frac{e^{-\lap n}}{n}\sin n\th_1\cos n\th_2=0,\\
\hf\frac{\del F}{\del\th_2}&=2\th_2-\pi+\bt_2\sin\th_2
+2\sum_{n=1}^\infty \frac{e^{-\lap n}}{n}\cos n\th_1\sin n\th_2=0,
\end{aligned} 
\label{eq:saddleeq-2pt}
\end{equation}
where the last terms in \eqref{eq:saddleeq-2pt}
are also written as
\begin{equation}
\begin{aligned}
 2\sum_{n=1}^\infty \frac{e^{-\lap n}}{n}\sin n\th_1\cos n\th_2&
=\arctan\left(\frac{\sin(\th_1+\th_2)}{e^{\lap}-\cos(\th_1+\th_2)}\right)
+\arctan\left(\frac{\sin(\th_1-\th_2)}{e^{\lap}-\cos(\th_1-\th_2)}\right),\\
 2\sum_{n=1}^\infty \frac{e^{-\lap n}}{n}\cos n\th_1\sin n\th_2&
=\arctan\left(\frac{\sin(\th_1+\th_2)}{e^{\lap}-\cos(\th_1+\th_2)}\right)
-\arctan\left(\frac{\sin(\th_1-\th_2)}{e^{\lap}-\cos(\th_1-\th_2)}\right).
\end{aligned} 
\end{equation}

As discussed in \cite{Goel:2023svz},
we can solve the saddle point equation \eqref{eq:saddleeq-2pt}
order by order in the small $\lap$ expansion. 
To this end, it is convenient to set
\begin{equation}
\begin{aligned}
 \bt_1=\frac{u-\phi}{\cos u},\quad \bt_2=\frac{u+\phi}{\cos u}.
\end{aligned} 
\end{equation}
From \eqref{eq:bt-tau}, one can see that
\begin{equation}
\begin{aligned}
 \bt=\bt_1+\bt_2=\frac{2u}{\cos u},\quad \phi=\left(1-\frac{2\tau_{12}}{\bt}\right)u.
\end{aligned} 
\label{eq:phi-def}
\end{equation}
The saddle point value of $\th_k~(k=1,2)$ is expanded as
\begin{equation}
\begin{aligned}
 \th_k=\frac{\pi}{2}-u+a_k\lap+b_k\lap^2+\cO(\lap^3).
\end{aligned} 
\label{eq:thk-saddle}
\end{equation}
As discussed in \cite{Goel:2023svz},
the leading term of \eqref{eq:thk-saddle}
is the same as the saddle point for the partition function
\eqref{eq:th*}.
At the $\cO(\lap^0)$ of saddle point equation, we find
\begin{equation}
\begin{aligned}
 a_1-a_2=\tan\phi.
\end{aligned} 
\end{equation}
At the next order $\cO(\lap)$ of saddle-point equation, we find
\begin{equation}
\begin{aligned}
 a_1+a_2&=\frac{(1+\phi\tan\phi)\tan u}{1+u\tan u},\\
b_1-b_2
&=\frac{1}{2\cos^2\phi}\left[-(1+u\tan u)\tan\phi
+\frac{\phi\tan^2u (1+\phi\tan\phi)}{1+u\tan u}\right].
\end{aligned} 
\end{equation}

Let us compute the saddle point value of $F$. We are interested in the regime where
$\la$ and $\lap$ are of the same order
\begin{equation}
\begin{aligned}
 \lap\sim\cO(\la),\quad\frac{\lap}{\la}=\text{finite}.
\end{aligned} 
\end{equation}
Then, in order to evaluate $\til{G}_2$ up to $\cO(\la)$, we have to
compute the saddle point value of $F$ up to $\cO(\lap^2)$ since
\begin{equation}
\begin{aligned}
 \frac{\lap^2}{\la}\sim\cO(\la).
\end{aligned} 
\label{eq:scaling}
\end{equation}
We expand the saddle point value of $F$ as
\begin{equation}
\begin{aligned}
 -F_*=f_0+f_1\lap+\hf I\lap^2 +\cO(\lap^3).
\end{aligned} 
\label{eq:F-saddle}
\end{equation}
One can compute $f_1$ and $I$ by using the following relation
\begin{equation}
\begin{aligned}
 \frac{dF_*}{d\lap}&=\frac{\del F_*}{\del\lap}+\sum_{k=1,2}
\frac{\del\th_k}{\del\lap}\frac{\del F_*}{\del\th_k}\\
&=\frac{\del F_*}{\del\lap},
\end{aligned} 
\end{equation}
where  in the last equality we used the saddle point equation \eqref{eq:saddleeq-2pt}.
From \eqref{eq:Fh-2pt} we find
\begin{equation}
\begin{aligned}
\frac{dF_*}{d\lap}=\log\frac{\cosh^2\lap}{\bigl[\cosh\lap-\cos(\th_1+\th_2)\bigr]
\bigl[\cosh\lap-\cos(\th_1-\th_2)\bigr]}. 
\end{aligned} 
\label{eq:dFdlap}
\end{equation}
Plugging the saddle point value \eqref{eq:thk-saddle}
into \eqref{eq:dFdlap} and expanding in $\lap$, we find
that $f_1$ and $I$ in \eqref{eq:F-saddle} are given by
\begin{equation}
\begin{aligned}
 f_1=\log\frac{\cos^2u}{\cos^2\phi},\quad I=\frac{f(\phi)f(-\phi)}{1+u\tan u},
\end{aligned} 
\label{eq:I-def}
\end{equation}
where $f(x)$ is defined by
\begin{equation}
\begin{aligned}
 f(x)=(1+u\tan u)\tan x-(1+x\tan x)\tan u.
\end{aligned} 
\label{eq:fx}
\end{equation}
The first term $f_0$ in \eqref{eq:F-saddle} is equal to the leading order free energy \eqref{eq:f0} and it is canceled when we normalize by the partition function
\eqref{eq:normalizedG}. 
At the one-loop level, we have to perform the
Gaussian integral around the saddle point
and evaluate the one-loop determinant
\begin{equation}
\begin{aligned}
 \frac{1}{\rt{\det F_{ij}}},
\end{aligned} 
\end{equation} 
where $F_{ij}=\frac{\del^2F}{\del\th_i\del\th_j}$ is the Hessian at the saddle point.
This computation was already carried out in \cite{Goel:2023svz} and we will not repeat
it here.\footnote{As we discuss in appendix \ref{app:alt},
the Hessian can be diagonalized by a change of variables.}
Finally, we find the normalized two-point function 
\eqref{eq:normalizedG} at the one-loop order
\begin{equation}
\begin{aligned}
 G_2
&=\left(\frac{\cos^2u}{\cos^2\phi}\right)^{\frac{\lap}{\la}}
\Biggl(1+\frac{\lap^2}{2\la}I+\lap\cA\Biggr)\\
&\approx e^{\frac{\lap^2}{2\la}I}\left[\frac{\cos^2u}{\cos^2\phi}(1+\la \cA)\right]^{\frac{\lap}{\la}},
\end{aligned} 
\label{eq:2pt-1loop}
\end{equation}
where $\cA$ is given by
\begin{equation}
\begin{aligned}
 \cA
&=\frac{1}{4(1+u\tan u)}\Biggl[-\frac{(1+u\tan u)^2}{\cos^2\phi}+\frac{(1+\phi\tan \phi)^2}{\cos^2u}+\phi^2(\tan^2u-\tan^2\phi)-\frac{1+\phi\tan\phi}{1+u\tan u}+1\Biggr].
\end{aligned} 
\label{eq:A-2pt}
\end{equation}
We have checked that our $\cA$ agrees with $G_1$ in \cite{Goel:2023svz}.
One can easily see that 
\begin{equation}
\begin{aligned}
 I|_{\phi=\pm u}=\cA|_{\phi=\pm u}=0.
\end{aligned} 
\end{equation}
Our key observation is that $\cA$ in \eqref{eq:A-2pt}
satisfies a simple relation
\begin{equation}
\begin{aligned}
 \left(\del_\phi^2-\frac{2}{\cos^2\phi}\right)\cA=\frac{1}{\cos^2\phi}I.
\end{aligned} 
\label{eq:A-del}
\end{equation}
In section \ref{sec:Liouville}, we will see that this relation naturally 
follows from the computation in the Liouville theory.
We note in passing that $f(x)$ in \eqref{eq:fx} satisfies
\begin{equation}
\begin{aligned}
 f(u)=0,\qquad 
\left(\del_x^2-\frac{2}{\cos^2x}\right)f(x)=0 \, ,
\end{aligned} 
\end{equation}
as this is the zero-mode wave function of the Liouville theory as we explain in the next section.

\subsection{Uncrossed four-point function}
Next, let us consider the small $\la$
expansion of the uncrossed four-point function
\begin{equation}
\begin{aligned}
 \til{G}_4&=\int\prod_{i=1}^3\frac{d\th_i}{2\pi}\mu(\th_i)e^{-\bt_i E(\th_i)}
\frac{(e^{-2\lap_1};q)_\infty}{(e^{-\lap_1+\ri(\pm\th_1\pm\th_3)};q)_\infty}
\frac{(e^{-2\lap_2};q)_\infty}{(e^{-\lap_2+\ri(\pm\th_2\pm\th_3)};q)_\infty}\\
&=
\int\prod_{i=1}^3\frac{d\th_i}{2\pi}e^{-\la^{-1}F+h+\cO(\la)},
\end{aligned} 
\end{equation}
where
\begin{equation}
\begin{aligned}
 F&=\sum_{i=1}^3\Biggl[2\left(\th_i-\frac{\pi}{2}\right)^2-2\bt_i\cos\th_i\Biggr]
+\text{Li}_2(e^{-2\lap_1})+\text{Li}_2(e^{-2\lap_2})\\
&\qquad 
-\text{Li}_2(e^{-\lap_1+\ri(\pm\th_1\pm\th_3)})
-\text{Li}_2(e^{-\lap_2+\ri(\pm\th_2\pm\th_3)}),\\
h&=\sum_{i=1}^3\Biggl[\hf\bt_i\cos\th_i+\log(2\sin\th_i)\Biggr]
+\hf\log(1-e^{-2\lap_1})+\hf\log(1-e^{-2\lap_2})\\
&\qquad-\hf\log(1-e^{-\lap_1+\ri(\pm\th_1\pm\th_3)})
-\hf\log(1-e^{-\lap_2+\ri(\pm\th_2\pm\th_3)}).
\end{aligned} 
\end{equation}
The saddle point equation reads
\begin{equation}
\begin{aligned}
 \hf\frac{\del F}{\del\th_1}&=2\th_1-\pi+\bt_1\sin\th_1
+\arcsin\left(\frac{\sin(\th_1+\th_3)}{e^{\lap_1}-\cos(\th_1+\th_3)}\right)
+\arcsin\left(\frac{\sin(\th_1-\th_3)}{e^{\lap_1}-\cos(\th_1-\th_3)}\right),\\
\hf\frac{\del F}{\del\th_2}&=2\th_2-\pi+\bt_2\sin\th_2
+\arcsin\left(\frac{\sin(\th_2+\th_3)}{e^{\lap_2}-\cos(\th_2+\th_3)}\right)
+\arcsin\left(\frac{\sin(\th_2-\th_3)}{e^{\lap_2}-\cos(\th_2-\th_3)}\right),\\
\hf\frac{\del F}{\del\th_3}&=2\th_3-\pi+\bt_3\sin\th_3
+\arcsin\left(\frac{\sin(\th_1+\th_3)}{e^{\lap_1}-\cos(\th_1+\th_3)}\right)
-\arcsin\left(\frac{\sin(\th_1-\th_3)}{e^{\lap_1}-\cos(\th_1-\th_3)}\right)\\
&\qquad+\arcsin\left(\frac{\sin(\th_2+\th_3)}{e^{\lap_2}-\cos(\th_2+\th_3)}\right)
-\arcsin\left(\frac{\sin(\th_2-\th_3)}{e^{\lap_2}-\cos(\th_2-\th_3)}\right).
\end{aligned} 
\label{eq:saddleeq-4pt}
\end{equation}
As in the previous subsection, one can solve this saddle point equation
order by order in the small $\lap_i$ expansion.
To this end, it is convenient to parameterize $\bt_{i}~(i=1,2,3)$ as
\begin{equation}
\begin{aligned}
 \bt_1=\frac{u-\phi_1}{\cos u},\quad
\bt_2=\frac{u-\phi_2}{\cos u},\quad
\bt_3=\frac{\phi_1+\phi_2}{\cos u}.
\end{aligned} 
\label{eq:def-phi12}
\end{equation}
Then the saddle point solution is expanded as
\begin{equation}
\begin{aligned}
 \th_1&=\frac{\pi}{2}-u+\frac{a_1+\tan\phi_1}{2}\lap_1
+\frac{a_2+\tan\phi_2}{2}\lap_2+b_1\lap_1^2
+b_2\lap_1\lap_2,\\
\th_2&=\frac{\pi}{2}-u+\frac{a_1+\tan\phi_1}{2}\lap_1
+\frac{a_2+\tan\phi_2}{2}\lap_2+c_1\lap_2^2
+c_2\lap_1\lap_2,\\
\th_3&=\frac{\pi}{2}-u
+\frac{a_1-\tan\phi_1}{2}\lap_1
+\frac{a_2-\tan\phi_2}{2}\lap_2,
\end{aligned} 
\end{equation}
where
\begin{equation}
\begin{aligned}
 a_1&=\frac{(1+\phi_1\tan\phi_1)\tan u}{1+u\tan u},\\
b_1&=\frac{1}{2\cos^2\phi_1}\left[
-(1+u\tan u)\tan\phi_1+\frac{\phi_1(1+\phi_1\tan\phi_1)\tan^2u}{1+u\tan u}
\right],\\
b_2&=\frac{1}{2\cos^2\phi_1}\left[
(1+u\tan u)\tan\phi_2+\frac{\phi_1(1+\phi_2\tan\phi_2)\tan^2u}{1+u\tan u}
-(1+\phi_2\tan\phi_2+\phi_1\tan\phi_2)\tan u
\right],\\
a_2&=a_1\Big|_{\phi_1\leftrightarrow\phi_2},\quad
c_1=b_1\Big|_{\phi_1\leftrightarrow\phi_2},\quad
c_2=b_2\Big|_{\phi_1\leftrightarrow\phi_2}.
\end{aligned} 
\end{equation}
We also expand the saddle point value of $F$ up to $\cO(\lap^2)$
\begin{equation}
\begin{aligned}
-F_*=f_0+\sum_{i=1,2}f_i\lap_i+\hf\sum_{i,j=1,2}
I_{ij}\lap_i\lap_j.
\end{aligned} 
\end{equation}
Here $f_0$ is the leading order free energy \eqref{eq:f0} and 
$f_i=\log\frac{\cos^2u}{\cos^2\phi_i}~(i=1,2)$. We find that the diagonal part of $I_{ij}$ is equal to the corresponding $I$
in the two-point function \eqref{eq:I-def}
\begin{equation}
\begin{aligned}
 I_{ii}=I\big|_{\phi=\phi_i}=\frac{f(\phi_i)f(-\phi_i)}{1+u\tan u}, \quad (i=1,2),
\end{aligned} 
\end{equation}
and the off-diagonal part $I_{12}$ is given by
\begin{equation}
\begin{aligned}
 I_{12}=\frac{f(\phi_1)f(\phi_2)}{1+u\tan u},
\end{aligned}
\label{eq:I12} 
\end{equation}
where $f(x)$ is defined in \eqref{eq:fx}.
The computation of the one-loop determinant $\det F_{ij}$ is almost parallel to 
the two-point function.
After some algebra, 
we find the uncrossed four-point function at the one-loop level
\begin{equation}
\begin{aligned}
 G_4=e^{\frac{1}{2\la}I_{ij}\lap_i\lap_j}\prod_{i=1,2}\left[\frac{\cos^2u}{\cos^2\phi_i}(1+\la \cA_i)\right]^{\frac{\lap_i}{\la}},
\end{aligned}
\label{eq:4pt-1loop} 
\end{equation}
where $\cA_i$ is the same as the one-loop correction
$\cA$ \eqref{eq:A-2pt}
appeared in the two-point function
\begin{equation}
\begin{aligned}
 \cA_i=\cA|_{\phi=\phi_i}.
\end{aligned} 
\end{equation}
Our result \eqref{eq:4pt-1loop} 
is a generalization of the one-loop computation of the two-point function
in \cite{Goel:2023svz}.
We should stress that the factor $\frac{1}{2\la}I_{ij}\lap_i\lap_j$ was not considered in 
\cite{Goel:2023svz}, but it should be included in the scaling regime 
\eqref{eq:scaling}.

\subsubsection{Relation to the energy fluctuation}
If we normalize the four-point function \eqref{eq:4pt-1loop} 
by the two-point function \eqref{eq:2pt-1loop} at the one-loop level, we find
\begin{equation}
\begin{aligned}
 \frac{G_4}{G_2(\phi_1)G_2(\phi_2)}=e^{\frac{\lap_1\lap_2}{\la}I_{12}}.
\end{aligned} 
\label{eq:I12-int}
\end{equation}
This relation suggests that $I_{12}$ can be thought of as the interaction term of the two
particles in the bulk spacetime corresponding to the boundary operators
with dimension $\lap_1$ and $\lap_2$.
As discussed in \cite{Maldacena:2016hyu}, this interaction
can be understood from the coupling of the matter operator
to the energy fluctuation
$\h{H}=H-\bra H\ket$.
Let us repeat the argument in \cite{Maldacena:2016hyu}. 
The saddle-point value of the energy is
\begin{equation}
\begin{aligned}
 E=-\frac{2\cos\th_*}{\la}=-\frac{2\sin u}{\la},
\end{aligned} 
\end{equation}
where we used \eqref{eq:th*}.
Thus the energy fluctuation is related to $\cob u$ by
\begin{equation}
\begin{aligned}
 \cob E=-\frac{2\cos u}{\la}\cob u.
\end{aligned} 
\label{eq:E-u}
\end{equation}
Using the relation
\begin{equation}
\begin{aligned}
 \phi=\left(1-\frac{2\tau}{\bt}\right)u,
\end{aligned} 
\end{equation}
the fluctuation of $\phi$ under the variation of
$\bt$ is written as
\begin{equation}
\begin{aligned}
 \cob \phi&=\frac{2\tau u}{\bt}\frac{\cob \bt}{\bt}+\phi \frac{\cob u}{u}\\
&=(1+u\tan u-\phi\tan u)\cob u.
\end{aligned} 
\end{equation}
Here we have used
\begin{equation}
\begin{aligned}
 \frac{\cob \bt}{\bt}=\cob\log\bt=\cob\log\frac{2u}{\cos u}=(1+u\tan u)\frac{\cob u}{u}.
\end{aligned} 
\end{equation}
Then the change of two-point function is
\begin{equation}
\begin{aligned}
 \cob\log G_2&=\cob \log\left(\frac{\cos^2u}{\cos^2\phi}\right)^{\frac{\lap}{\la}}\\
&=\frac{2\lap}{\la}(\tan u\cob u-\tan\phi\cob\phi)\\
&=-\frac{2\lap}{\la}f(\phi)\cob u,
\end{aligned} 
\label{eq:cobG2}
\end{equation}
where $f(x)$ is defined in \eqref{eq:fx}.
From \eqref{eq:E-u}, the variance of $\cob u$ is estimated as
\begin{equation}
\begin{aligned}
 \bra(\cob u)^2\ket&=\frac{\la^2}{4\cos^2 u}\bra(\cob E)^2\ket\\
&=\frac{\la^2}{4\cos^2 u}\del_\bt^2\log Z.
\end{aligned} 
\label{eq:var-u}
\end{equation}
Plugging the leading order free energy
\begin{equation}
\begin{aligned}
 \log Z=\frac{2}{\la}(-u^2+2u\tan u)
\end{aligned} 
\end{equation}
into \eqref{eq:var-u}, we find
\begin{equation}
\begin{aligned}
 \bra(\cob u)^2\ket=\frac{\la}{4}\frac{1}{1+u\tan u}.
\end{aligned} 
\label{eq:u-variance}
\end{equation}
Finally, combining  \eqref{eq:cobG2} and \eqref{eq:u-variance} we find
\begin{equation}
\begin{aligned}
 \big\bra \cob\log G_2(\phi_1)\cob\log G_2(\phi_2)\ket=
\frac{\lap_1\lap_2}{\la}\frac{f(\phi_1)f(\phi_2)}{1+u\tan u}
=\frac{\lap_1\lap_2}{\la}I_{12}.
\end{aligned} 
\end{equation}
This precisely matches the interaction we found in
\eqref{eq:I12-int}.
This result implies that the off-diagonal part $I_{12}$ represents a total energy exchange of the external operators.
Note that $\cob u$ corresponds to $\rt{\la}\vep$ in \eqref{eq:vep2}
and the variance $\bra(\cob u)^2\ket$
in \eqref{eq:u-variance} agrees with the one 
obtained from the quadratic action for $\vep$ in \eqref{eq:vep2}.

\section{One-loop correction from Liouville theory}
\label{sec:Liouville}
In this section, we will show that the one-loop correction to the two- and four-point functions obtained in the previous section can be reproduced from
the Liouville theory.
As shown in \cite{Cotler:2016fpe},
the double scaling limit of the $G\Si$ action of the SYK model
reduces to the Liouville action for the bi-local field
$g(\tau_1,\tau_2)$
\begin{equation}
\begin{aligned}
 S&=\frac{1}{8\la}\int d\tau d\nu
\left[-\hf(\del_{\tau} g)^2+\hf(\del_{\nu} g)^2-2e^{g}\right],
\end{aligned}
\label{eq:Liouville-S} 
\end{equation}
where 
\begin{equation}
\begin{aligned}
 \tau=\tau_1-\tau_2,\quad \nu=\tau_1+\tau_2.
\end{aligned} 
\end{equation}
Introducing the coordinate $x,y$ by
\begin{equation}
\begin{aligned}
 x=u-\tau\cos u,\quad y=\nu\cos u,
\end{aligned} 
\end{equation}
the Liouville action is written as
\begin{equation}
\begin{aligned}
 S&=\frac{1}{8\la}\int_{-u}^u dx\int_0^{4u}dy 
\left[-\hf(\del_xg)^2+\hf(\del_yg)^2-\frac{2}{\cos^2u}e^g\right].
\end{aligned} 
\label{eq:L-action}
\end{equation}
Note that $x$ corresponds to $\phi$ in \eqref{eq:phi-def}.
We assumed that $\tau_1$ and $\tau_2$ are ordered pair of points on the thermal circle $S^1_\bt$
\begin{equation}
\begin{aligned}
0<\tau_2<\tau_1<\bt.
\end{aligned} 
\end{equation}
Then the range of $\tau,\nu$ and $x,y$ are related by
\begin{equation}
\begin{aligned}
0<\tau<\bt,~0<\nu<2\bt\quad\Rightarrow\quad
-u<x<u,~0<y<4u,
\end{aligned} 
\end{equation}
where $\bt$ and $u$ are related by \eqref{eq:saddle-u}.

The equation of motion following from the action \eqref{eq:L-action}
is
\begin{equation}
\begin{aligned}
 \del_x^2g-\del_y^2g-\frac{2}{\cos^2u}e^g=0.
\end{aligned} 
\label{eq:eom}
\end{equation}
One can easily see that
\begin{equation}
\begin{aligned}
 g_{\text{cl}}(x) \, = \, \log \left(\frac{\cos^2u}{\cos^2x} \right) \, ,
\end{aligned} 
\label{eq:gcl}
\end{equation}
is a ``static'' (i.e. independent of $y$) classical solution of the equation of motion \eqref{eq:eom}, with boundary conditions $g(x=\pm u)=0$.
Let us consider the expansion of the Liouville action
\eqref{eq:L-action}
around the classical solution \eqref{eq:gcl}
\begin{equation}
\begin{aligned}
 g(x,y)=g_{\text{cl}}(x)+\rt{\la}\vep(x,y).
\end{aligned} 
\end{equation}
The classical action is given by
\begin{equation}
\begin{aligned}
 S_{\text{cl}}&=\frac{1}{8\la}\int_{-u}^u dx\int_0^{4u}dy
\left[-\hf(\del_xg_{\text{cl}})^2-\frac{2}{\cos^2u}e^{g_{\text{cl}}}\right]\\
&=\frac{2}{\la}(u^2-2u\tan u),
\end{aligned} 
\end{equation}
which agrees with the leading order free energy in \eqref{eq:f0}.
The quadratic part of the action of $\vep$ becomes
\begin{equation}
\begin{aligned}
 S_2=\frac{1}{8}\int dx dy \left[-\hf(\del_x \vep)^2+\hf(\del_y \vep)^2
-\frac{2}{\cos^2x}\frac{\vep^2}{2}\right].
\end{aligned} 
\end{equation}
Then the propagator of $\vep$ is defined by
\begin{equation}
\begin{aligned}
\frac{1}{8} \left(
\del_x^2-\del_y^2-\frac{2}{\cos^2x}\right) \bra \vep(x,y)\vep(x',y')\ket
&=\cob(x-x')\h{\cob}(y-y'),
\end{aligned} 
\end{equation}
where $\h{\cob}(y-y')$ is the periodically extended $\cob$-function
\begin{equation}
\begin{aligned}
\h{\cob}(y-y')=\sum_{m\in\mathbb{Z}}\cob(y-y'+4um)=
\sum_{n\in\mathbb{Z}}
\frac{e^{\ri\til{n}(y-y')}}{4u},
\end{aligned} 
\end{equation}
with $\til{n}=n/v$ (see \eqref{eq:u-v} for the relation between $u$ and $v$). Note that the translation symmetry of $x$-coordinate is spontaneously broken by choosing the classical background $g_{\text{cl}}$, but the $y$-coordinate remains periodic with periodicity $4u=2\pi v$.

The propagator is also written as
\begin{equation}
\begin{aligned}
 \bra \vep(x,y)\vep(x',y')\ket=\frac{2}{u}\sum_{n\in\mathbb{Z}}D_n(x,x')
e^{\ri\til{n}(y-y')},
\end{aligned} 
\label{eq:prop-sum}
\end{equation}
where $D_n(x,x')$ satisfies
\begin{equation}
\begin{aligned}
 \left[\del_x^2-\frac{2}{\cos^2 x}+\til{n}^2\right]
D_n(x,x')=\cob(x-x').
\end{aligned}
\label{eq:Dn-eq} 
\end{equation}
One can solve \eqref{eq:Dn-eq} under the boundary condition
\begin{equation}
\begin{aligned}
 D_n(u,x')=D_n(-u,x')=0,\qquad
D_n(x',x)=D_n(x,x')=D_n(-x,-x').
\end{aligned} 
\end{equation} 
As is well-known, the solution of
\eqref{eq:Dn-eq} can be constructed from the 
two independent solutions of the homogeneous equation
\begin{equation}
\begin{aligned}
 \left[\del_x^2-\frac{2}{\cos^2 x}+\til{n}^2\right]f_n(x)=0,
\end{aligned} 
\end{equation}
whose explicit form is easily obtained as
\begin{equation}
\begin{aligned}
 f_n^{(1)}=\cos(\til{n}x)\tan x-\til{n}\sin(\til{n}x),\quad
f_n^{(2)}=\sin(\til{n}x)\tan x+\til{n}\cos(\til{n}x).
\end{aligned} 
\end{equation}
We can take a linear combination 
of $f^{(1)}_n(x)$ and $f^{(2)}_n(x)$
so that $f_n(x)$ vanishes at $x=u$
\begin{equation}
\begin{aligned}
 f_n(x)=f^{(1)}_n(x)f^{(2)}_n(u)-f^{(1)}_n(u)f^{(2)}_n(x).
\end{aligned} 
\end{equation}
Then the propagator for $n\ne0$ is given by
\begin{equation}
\begin{aligned}
 D_n(x,x')=
\frac{\th(x-x')f_n(x)f_n(-x')+\th(x'-x)f_n(-x)f_n(x')}{\{f_n(x),f_n(-x)\}},
\end{aligned} 
\label{eq:Dn}
\end{equation}
where the denominator is the Wronskian 
\begin{equation}
\begin{aligned}
 \{f_n(x),f_n(-x)\}&=\del_xf_n(x)f_n(-x)-f_n(x)\del_xf_n(-x)\\
&=2(-1)^n\til{n}^2(\til{n}^2-1)\tan u.
\end{aligned} 
\end{equation}
For the zero mode, we have\footnote{The zero-mode propagator $D_0(x,x')$ has been considered in \cite{Tarnopolsky:2018env}.}
\begin{equation}
\begin{aligned}
 D_0(x,x')=-\frac{\th(x-x')f(x)f(-x')+\th(x'-x)f(-x)f(x')}{2\tan u(1+u\tan u)},
\end{aligned} 
\label{eq:D0}
\end{equation}
where $f(x)$ is defined in \eqref{eq:fx}.
Note that the zero-mode wavefunction $f(x)$ is formally related to the non-zero mode 
$f_n(x)$ as
\begin{equation}
\begin{aligned}
 f(x)=\lim_{n\to0}\frac{1}{\til{n}}f_n(x).
\end{aligned} 
\end{equation}

Plugging $D_0(x,x')$ \eqref{eq:D0} and
$D_n(x,x')$ \eqref{eq:Dn} into \eqref{eq:prop-sum}, we find
\begin{equation}
\begin{aligned}
 \bra \vep(x,y)\vep(x',y')\ket=\frac{1}{u\tan u}\left[
-\frac{f(x)f(-x')}{1+u\tan u}+
\sum_{|n|\geq1}\frac{(-1)^n f_n(x)f_n(-x')}{\til{n}^2(\til{n}^2-1)}
e^{\ri\til{n}(y-y')}\right].
\end{aligned} 
\label{eq:prop-defined} 
\end{equation}
Here we assumed $x>x'$.
The sum over $n$ can be performed 
using the formula in \eqref{eq:sum-nt} and we find
\begin{equation}
\begin{aligned}
\sum_{|n|\geq1}\frac{(-1)^n f_n(x)f_n(-x')}{\til{n}^2(\til{n}^2-1)}e^{\ri\til{n}(y-y')}
&=f(x)f(-x').
\end{aligned} 
\end{equation}
Finally, the propagator becomes
\begin{equation}
\begin{aligned}
 \bra \vep(x,y)\vep(x',y')\ket=\frac{f(x)f(-x')}{1+u\tan u},\quad(x>x').
\end{aligned}
\label{eq:prop-result} 
\end{equation}
Note that this propagator is $y$-independent;
the $y$-independence of the time-ordered
four-point function was also mentioned in \cite{Maldacena:2016hyu}.
Note also that \eqref{eq:prop-result}
is finite at the coincident point $(x,y)=(x',y')$ and hence
there is no need of the normal ordering to define
the bi-local operator $e^{\frac{\lap}{\la}g(x,y)}$ at the perturbative level.

\subsection{One-loop correction of two-point function}
Let us compute the one-loop correction to the two-point function
from the Liouville theory.
The two-point function is defined by
\begin{equation}
\begin{aligned}
 G_2=\Bigl\bra e^{\frac{\lap}{\la}g(\phi,y_0)}\Bigr\ket=\frac{1}{Z}\int \mathcal{D}g \,e^{-S} 
e^{\frac{\lap}{\la}g(\phi,y_0)},
\end{aligned} 
\end{equation}
where $S$ is the Liouville action \eqref{eq:L-action}.
$y_0$ is some reference point but the result is independent of $y_0$
as we will see below.
To compute the one-loop correction, we expand the action
around the classical solution \eqref{eq:gcl} as
\begin{equation}
\begin{aligned}
 S-S_{\text{cl}}=\sum_{n=2}^\infty S_n=S_2+S_{\text{int}}
\end{aligned} 
\end{equation}
where
\begin{equation}
\begin{aligned}
 S_2&=\frac{1}{16}\int dxdy \,\vep(x,y) K\vep(x,y),\qquad K=
\del_x^2-\del_y^2-\frac{2}{\cos^2x},\\
S_n&=-\frac{\la^{\frac{n}{2}-1}}{4}\int dx dy \frac{1}{\cos^2x}\frac{\vep(x,y)^n}{n!},
\quad(n\geq3),\\
S_{\text{int}}&=\sum_{n\geq3}S_n.
\end{aligned} 
\end{equation}
Then the two-point function is written as
\begin{equation}
\begin{aligned}
 G_2&=e^{\frac{\lap}{\la}g_{\text{cl}}(\phi)}
\Bigl\bra e^{\frac{\lap}{\rt{\la}}\vep(\phi,y_0)-S_{\text{int}}}\Bigr\ket_{S_2}
=\left(\frac{\cos^2u}{\cos^2\phi}\right)^{\frac{\lap}{\la}}
\Bigl\bra e^{\frac{\lap}{\rt{\la}}\vep(\phi,y_0)-S_{\text{int}}}\Bigr\ket_{S_2}
\end{aligned} 
\end{equation}
where $\bra\cdots\ket_{S_2}$ is defined by
\begin{equation}
\begin{aligned}
 \bra\cdots\ket_{S_2}=\int\mathcal{D}\vep (\cdots)e^{-S_2}.
\end{aligned} 
\end{equation}
At the one-loop level, we find
\begin{equation}
\begin{aligned}
 \Bigl\bra e^{\frac{\lap}{\rt{\la}}\vep(\phi,y_0)-S_{\text{int}}}\Bigr\ket_{S_2}&=
\left\bra \Biggl[1+\frac{\lap}{\rt{\la}}\vep(\phi,y_0)
+\frac{\lap^2}{2\la}\vep(\phi,y_0)^2+\cdots\Biggr]\Biggl[1-S_3+\cdots\Biggr]\right\ket_{S_2}\\
&=1+\frac{\lap^2}{2\la}\bra\vep(\phi,y_0)^2\ket
+\frac{\lap}{4}\left\bra\vep(\phi,y_0)\int dxdy \frac{1}{\cos^2x}\frac{\vep(x,y)^3}{3!}\right\ket_{S_2}+\cdots
\end{aligned} 
\end{equation}
The second term reproduces $I$ in \eqref{eq:I-def}
\begin{equation}
\begin{aligned}
 I=\bra\vep(\phi,y_0)^2\ket
=\frac{f(\phi)f(-\phi)}{1+u\tan u},
\end{aligned} 
\end{equation}
and the last term corresponds to $\cA$ in \eqref{eq:A-2pt}
\begin{equation}
\begin{aligned}
\cA&=\frac{1}{4}\left\bra\vep(\phi,y_0)\int dxdy \frac{1}{\cos^2x}\frac{\vep(x,y)^3}{3!}\right\ket_{S_2}\\
&=\frac{1}{8}\int dx dy\frac{1}{\cos^2x}\bra\vep(\phi,y_0)\vep(x,y)\ket
\bra\vep(x,y)^2\ket.
\end{aligned} 
\label{eq:A-Liouville}
\end{equation}
From the expression of the propagator \eqref{eq:prop-result},
one can show that $\cA$ is independent of $y_0$.
Using this property, one can show that $\cA$ in \eqref{eq:A-Liouville}
satisfies the same relation \eqref{eq:A-del}
as we found for the one-loop correction in the previous section
\begin{equation}
\begin{aligned}
 \left(\del_\phi^2-\frac{2}{\cos^2\phi}\right)\cA&=
\left(\del_\phi^2-\del_{y_0^2}-\frac{2}{\cos^2\phi}\right)\cA\\
&=\int dx dy\frac{1}{\cos^2x}\cob(x-\phi)\h{\cob}(y-y_0)
\bra\vep(x,y)^2\ket
\\
&=\frac{1}{\cos^2\phi}\bra\vep(\phi,y_0)^2\ket\\
&=\frac{1}{\cos^2\phi}I.
\end{aligned} 
\end{equation}
This indeed reproduces \eqref{eq:A-del}.

\subsection{Uncrossed four-point function}
Next, let us consider the uncrossed four-point function.
In the Liouville language, the uncrossed four-point function is given by
\begin{equation}
\begin{aligned}
 G_4(\phi_1,\phi_2)=\left\bra e^{\frac{\lap_1}{\la}g(\phi_1,y_1)}
e^{\frac{\lap_2}{\la}g(-\phi_2,y_2)}\right\ket.
\end{aligned} 
\label{eq:4pt-Liouville}
\end{equation}
We have changed the sign of $\phi$ for one of the bi-local operator 
$e^{\frac{\lap}{\la}g}$. 
Note that $\tau$ and $\phi$ are related by
\begin{equation}
\begin{aligned}
 \phi(\tau)=\left(1-\frac{2\tau}{\bt}\right)u,\quad
\phi(\bt-\tau)=-\phi(\tau),
\end{aligned} 
\end{equation}
and thus the sign flip of $\phi$ corresponds to $\tau\to\bt-\tau$.
The necessity of the sign flip $\phi_2\to-\phi_2$ 
in \eqref{eq:4pt-Liouville}
is understood from the following picture
\begin{equation}
\begin{aligned}
 \begin{tikzpicture}[scale=0.75]
\draw (0,0) circle [radius=2];
\draw[blue,thick] (-1.73,-1)--(1.73,-1); 
\draw[red,thick] (-1.73,1)--(1.73,1); 
\draw[red,fill=red] (-1.73,1) circle [radius=0.1];
\draw[red,fill=red] (1.73,1) circle [radius=0.1];
\draw[blue,fill=blue] (-1.73,-1) circle [radius=0.1];
\draw[blue,fill=blue] (1.73,-1) circle [radius=0.1];
\draw[stealth-stealth,dashed] (1.9,1.1) arc (30:150:2.2);
\draw (0,2.2) node [above]{$\textcolor{red}{\tau_{34}}$};
\draw[stealth-stealth] (1.9,1.1) arc (30:-210:2.2);
\draw (0,-2.4) node [below]{$\textcolor{blue}{\tau_{12}}$};
\draw[stealth-stealth,dashed] (2.08,-1.2) arc (-30:-150:2.4);
\draw (-3.5,-0.5) node [left]{$\textcolor{red}{\bt-\tau_{34}}$};
\draw[->] (-3.5,-0.4)--(-2.3,0);
\end{tikzpicture}
\end{aligned}\qquad.
\label{minus-phi}
\end{equation}
Namely, $\phi$ of the bi-local operator is defined
with respect to the Hartle-Hawking state  $|0\ket$ 
\cite{Lin:2022rbf,Okuyama:2022szh} at the bottom of the figure in 
\eqref{minus-phi} and we have to choose
\begin{equation}
\begin{aligned}
 \phi_1=\phi(\tau_{12}),\quad
\phi_2=\phi(\bt-\tau_{34})=-\phi(\tau_{34}).
\end{aligned} 
\end{equation}

One can easily generalize the perturbative computation 
in the previous subsection to
the four-point function
\begin{equation}
\begin{aligned}
 G_4&=e^{\frac{\lap_1}{\la}g_{\text{cl}}(\phi_1)}
e^{\frac{\lap_2}{\la}g_{\text{cl}}(-\phi_2)}
\left\bra e^{\frac{\lap_1}{\rt{\la}}\vep(\phi_1,y_1)+
\frac{\lap_2}{\rt{\la}}\vep(-\phi_2,y_2)-S_{\text{int}}}\right\ket_{S_2}\\
&=\prod_{i=1,2}\left(\frac{\cos^2u}{\cos^2\phi_i}\right)^{\frac{\lap_i}{\la}}
\Biggl\bra
1+\frac{\lap_1^2}{2\la}\vep(\phi_1,y_1)^2
+\frac{\lap_2^2}{2\la}\vep(-\phi_2,y_2)^2+
\frac{\lap_1\lap_2}{\la}\vep(\phi_1,y_1)\vep(-\phi_2,y_2)\\
&\hskip40mm-\left(\frac{\lap_1}{\rt{\la}}\vep(\phi_1,y_1)+
\frac{\lap_2}{\rt{\la}}\vep(-\phi_2,y_2)\right)S_3+\cO(\la^2)
\Biggr\ket_{S_2}\\
&=e^{\frac{1}{2\la}I_{ij}\lap_i\lap_j}\prod_{i=1,2}
\left[\frac{\cos^2u}{\cos^2\phi_i}(1+\la \cA_i)\right]^{\frac{\lap_i}{\la}}.
\end{aligned} 
\label{eq:G4-pert}
\end{equation}
Using the explicit form of the propagator of
$\vep(x,y)$ in \eqref{eq:prop-result}, 
one can see that this computation reproduces the result of four-point function in the 
previous section.
For instance, from \eqref{eq:G4-pert} one can read off $I_{12}$ as
\begin{equation}
\begin{aligned}
I_{12}=\bra\vep(\phi_1,y_1)\vep(-\phi_2,y_2)\ket=
\frac{f(\phi_1)f(\phi_2)}{1+u\tan u},
\end{aligned} 
\end{equation}
which reproduces $I_{12}$ in \eqref{eq:I12} obtained from the saddle point analysis.
One can also show that $I_{ii},\cA_i~(i=1,2)$ are reproduced from \eqref{eq:G4-pert}
in a similar manner.

\section{Out-of-time-order correlators}
\label{sec:OTOC}
In this section, we study direction evaluation of the summation over $n$ in (\ref{eq:prop-defined}) for the out-of-time-ordered case: $\tau_1>\tau_3>\tau_2>\tau_4$.

Let us first consider a special case with $\tau_3 = \pi$ and $\tau_4=0$ as in \cite{Maldacena:2016hyu}, where we used $\beta=2\pi$ unit.
This corresponds to $x'=0$ and $y'=u$, as well as $\pi<\tau_1<2\pi$ and $0<\tau_2<\pi$.
In this case, one of the wave function is reduced to 
	\begin{align}
		f_n(0) \, = \, - \til{n} \, f_n^{(1)}(u) \, ,
	\end{align}
so that the Fourier series of the non-zero modes is rewritten as
	\begin{align}
		\sum_{|n|\ge1} D_n(x\, ,x') \, e^{\ri\til{n}(y-y')} \, = \, \frac{\cot u}{2} \sum_{|n|\ge1} \frac{(-1)^{n+1}}{\til{n} (\til{n}^2 - 1)} \,
        f_n(x) f_n^{(1)}(u) \, e^{\ri \til{n} y} \, e^{-\frac{\ri n\pi}{2}} \, .
	\end{align}
The summation over $n$ can be explicitly performed by using the formulae (\ref{formu1}) - (\ref{formu3}) as
	\begin{align}
		\sum_{|n|\ge1} D_n(x\, ,x') \, e^{\ri\til{n}(y-y')}
        \, = \, \frac{1}{2} \left[ \, - \tan x + (1+x \tan x)\tan u \, - \, u \, \frac{\cos(2u-y)}{\cos u \cos x} \right] \, .
	\end{align}
Finally combining with the zero mode contribution (\ref{eq:D0}), the out of time ordered two-point function is found as
	\begin{align}
		\langle \vep(x,y) \vep(0,u) \rangle \, = \, \frac{\tan^2 u (1+x \tan x)}{1+u \tan u} \, - \, \frac{\cos(2u-y)}{\cos u \cos x} \, .
	\end{align}	
This agrees with the results found in \cite{Streicher:2019wek, Choi:2019bmd}.
One can also check that the low temperature limit of this result agrees with the one found in \cite{Maldacena:2016hyu} (see section~\ref{sec:low}).

In order to obtain the Lyapunov exponent, we set $x=0$ and $y=-\ri 2\pi v t /\beta$, which gives
	\begin{align}
		\sum_{|n|\ge1} D_n(x\, ;x') \, e^{\ri \til{n}(y-y')} \, = \, - \, \frac{u\cos u}{2} \, \cosh \left( \frac{2\pi v t}{\beta} \right) \, + \, \cdots \, .
	\end{align}
From this we find 
	\begin{align}
		\la_L \, = \,  \frac{2\pi v}{\beta} \, , 
	\end{align}
which agrees with the Lyapunov exponent found in \cite{Maldacena:2016hyu}.

Next, let us try produce the $x'$-dependence as well.
For this purpose, we keep $x'$ general (but assume close to $0$) and set $y'=u$.
We also assume $\pi<\tau_1<2\pi$ and $0<\tau_2<\pi$ and use the formulae (\ref{formu1}) - (\ref{formu3}).
This leads to 
	\begin{align}
		\sum_{|n|\ge1} D_n(x\, ,x') \, e^{\ri\til{n}(y-y')} \, &= \, \frac{1}{2} \bigg[ \tan u \, f^{(2)}(x) f^{(2)}(x') \, + \, \tan x' f^{(2)}(x) \, - \, \tan x \, f^{(2)}(x') \nn\\
		&\quad \, - \, \frac{u \cos(2u-y)}{\cos u \cos x \cos x'} \, + \, \big( (y-2u) u \tan u - (u+\cot u) \big)  \tan x \tan x' \bigg] \, , 
 	\end{align}
where $f^{(2)}(x) = 1 + x \tan x$.
Finally combining with the zero mode contribution (\ref{eq:D0}), the out of time ordered two-point function is found as
	\begin{align}
		\langle \vep(x,y) \vep(x',y') \rangle \, &= \, \frac{\tan^2 u}{1+u \tan u} \, f^{(2)}(x) f^{(2)}(x') \\
        &\qquad \, - \, \frac{\cos(2u-y)}{\cos u \cos x \cos x'} \, + \,(y-2u) \tan u \tan x \tan x' \, . \nn
	\end{align}
this result completely agrees with the results found in \cite{Streicher:2019wek, Choi:2019bmd}.

\section{Low temperature limit}
\label{sec:low}
In this section, we will give low temperature expressions of our results obtained above, and compare with previous works \cite{Maldacena:2016hyu}.

Let us start from the two-point function. For the function $\mathcal{A}$ defined in (\ref{eq:A-2pt}),
using
	\begin{align}
		u \, = \, \frac{\pi}{2} \, - \, \frac{\pi}{\beta} \, , \qquad \phi \, = \, \frac{\pi}{2} \, - \, \frac{\pi \tau_{12}}{\beta} \, ,
	\end{align}
the low temperature limit of $\mathcal{A}$ is found as
	\begin{align}
		\mathcal{A} \, = \, \frac{\beta}{2\pi^2} \left[ 1 \, - \, \frac{\frac{\pi^2 \tau_{12}}{\beta} (1 - \frac{\tau_{12}}{\beta})}{\sin^2 (\frac{\pi \tau_{12}}{\beta})} \, + \, \frac{\pi(1 - \frac{2\tau_{12}}{\beta})}{\tan\frac{\pi \tau_{12}}{\beta}} \right] \, .
	\end{align}
Up to an overall coefficient, this agrees with $\langle \mathcal{C}(u_1, u_2)\rangle$ computed in Schwarzian theory in the next subsection~\ref{sec:schwarzian_2pt}.
The low temperature limit of the function $f(x)$ defined in (\ref{eq:fx}) is given by
	\begin{align}
		f(x) \, = \, - \, \frac{\beta}{\pi} \left( 1 \, - \, \frac{\pi \tau_{12}}{\beta} \cot \frac{\pi \tau_{12}}{\beta} \right) \, .
	\end{align}
This guarantees that the four-point function in the low temperature limit agrees with the low temperature result found in \cite{Maldacena:2016hyu}.
This also shows that the low temperature limit of the function $I$ defined in (\ref{eq:I-def}) agrees with $\langle \mathcal{B}(u_1, u_2) \mathcal{B}(u_3, u_4) \rangle$
computed in Schwarzian theory, up to an overall coefficient.

In appendix~\ref{app:factorization}, we will also show that the zero-temperature factorization holds at arbitrary $\lambda$ in the DSSYK model.

\subsection{One-loop correction from Schawrzian mode}
\label{sec:schwarzian_2pt}
In this subsection, we study the one-loop correction of two-point functions in Schwarzian theory.
Here we follow the notation of \cite{Maldacena:2016upp} and in particular $u_i=2\pi \tau_i/\beta$.
The reparametrization symmetry of two-point function transforms
	\begin{align}
		G^{(2)}_\Delta(\tau_1, \tau_2) \, = \, \frac{1}{|\tau_{12}|^{2\Delta}} \, \Rightarrow \, \left( \frac{f'(u_1) f'(u_2)}{(f(u_1) - f(u_2))^2} \right)^\Delta \, ,
 	\end{align} 
where $\tau_{ij}:=\tau_i - \tau_j$.
Parametrizing $f(u) = \tan((u + \vep(u))/2)$ and expanding up to $\vep^2$ order, we find
	\begin{align}
		G^{(2)}_\Delta(\tau_1, \tau_2) \, = \, \frac{1}{(2\sin\frac{u_{12}}{2})^{2\Delta}} \left[ 1 \, + \, \Delta \, \mathcal{B}(u_1, u_2) \, + \, \frac{\Delta^2}{2} \, \mathcal{B}(u_1, u_2)^2 \, + \, \Delta \, \mathcal{C}(u_1, u_2) \, + \, \cdots \right] \, ,
 	\label{G2_rep}
	\end{align} 
where
	\begin{align}
		\mathcal{B}(u_1, u_2) \, &= \, \vep'(u_1) \, + \, \vep'(u_2) \, - \, \frac{\vep(u_1) - \vep(u_2)}{\tan \frac{u_{12}}{2}} \, , \\
		\mathcal{C}(u_1, u_2) \, &= \, - \, \left( \frac{\vep'(u_1)^2 + \vep'(u_2)^2}{2} \right) \, + \, \frac{\big(\vep(u_1) - \vep(u_2) \big)^2}{4\sin^2( \frac{u_{12}}{2})} \, .
 	\end{align} 
Now we would like to evaluate the expectation value of the RHS of (\ref{G2_rep}) with using Schwarzian mode propagator
	\begin{align}
		\big\langle \vep(u_1) \vep(u_2) \big\rangle \, = \, \frac{1}{2\pi C} \left[ - \frac{(|u_{12}|-\pi)^2}{2} \, + \, (|u_{12}| - \pi) \sin |u_{12}| + a + b \cos u_{12} \right] \, ,
 	\end{align} 
where $C$ is the Schwarzian coupling and $a$ and $b$ are gauge parameters which should not appear in any physical quantities.
Since the one-point function of the Schwarzian mode vanishes, we have
	\begin{align}
		\big\langle \mathcal{B}(u_1, u_2) \big\rangle \, = \, 0 \, .
 	\end{align} 
The two-point function of $\mathcal{B}$ is evaluated in \cite{Maldacena:2016upp} as
	\begin{align}
		\big\langle \mathcal{B}(u_1, u_2) \mathcal{B}(u_3, u_4) \big\rangle
  \, = \, \frac{1}{2\pi C} \left( -2 + \frac{u_{12}}{\tan \frac{u_{12}}{2}} \right) \left( -2 + \frac{u_{34}}{\tan \frac{u_{34}}{2}} \right) \, .
 	\end{align} 
Finally we can also evaluate the one-point function of $\mathcal{C}$ as
	\begin{align}
		\big\langle \mathcal{C}(u_1, u_2) \big\rangle \, = \, \frac{1}{2\pi C} \left[ 1 \, + \, \frac{u_{12}(u_{12} - 2\pi)}{4\sin^2 (\frac{u_{12}}{2})} \, + \, \frac{(\pi - u_{12})}{\tan\frac{u_{12}}{2}} \right] \, .
 	\end{align}

\section{Conclusions and outlook}
\label{sec:conclusion}
In this paper, we have studied the one-loop correction to the correlators of DSSYK
from two approaches: the saddle point approximation of the exact result obtained from the chord diagrams, and the perturbative computation in the 
Liouville theory.
We found that the 
relation \eqref{eq:A-del} obeyed by the one-loop correction $\cA$ 
naturally follows from the  computation in the Liouville theory.
In particular, $\cA$ and $I_{ij}$ are closely related to the propagator
of the fluctuation $\vep(x,y)$ around the classical solution 
$g_{\text{cl}}$ in the Liouville theory.
We also found that the out-of-time-order propagator
$\bra\vep(x,y)\vep(x',y')\ket$ in the Liouville theory correctly reproduces the
known result of OTOC in the literature \cite{Maldacena:2016hyu,Streicher:2019wek,Choi:2019bmd}.
We have also seen that the low temperature limit of the one-loop correction
$\cA$ is reproduced from the corresponding computation in the Schwarzian theory.

There are many interesting open questions.
The Liouville field $g(\tau_1,\tau_2)$ can be thought of as a quantum analogue of the bulk geodesic length between the two points $\tau_1,\tau_2$ on the boundary. The classical solution $g_{\text{cl}}$ corresponds to the geodesic length in the semi-classical bulk geometry and $\vep(x,y)$ represents its quantum fluctuation. It would be interesting to ``decode'' the bulk quantum geometry defined by the Liouville field $g(\tau_1,\tau_2)$ along the lines of \cite{Goel:2023svz}.

Our analysis was restricted to the small $\la,\lap$ regime. It would be 
interesting to generalize our analysis to the finite $\la,\lap$ case.
When $\lap$ becomes large, the corresponding matter operator is called the ``heavy operator''. It is expected that the insertion of heavy operator 
strongly back-reacts to the bulk geometry and the spacetime is pinched in 
the limit $\lap\to\infty$ \cite{Berkooz:2020uly}.
It would be interesting to understand the bulk gravitational 
interpretation of this phenomenon.

It is also important to understand the symmetry underlying the DSSYK.
In particular, it would be interesting to understand the
quantum group symmetry of DSSYK and its bulk interpretation.\footnote{It is curious that the $q$-oscillator representation of the transfer matrix of DSSYK in \cite{Berkooz:2018jqr} also appears in a statistical mechanical problem known as the Asymmetric Simple Exclusion Process (ASEP) \cite{blythe2000exact}. In this context, the quantum group symmetry naturally arises after mapping the problem of ASEP to the matrix product states of XXZ spin chain \cite{Crampe:2014aoa}.}
At finite $\la$, it is suggested that the bulk spacetime is discretized \cite{Lin:2022rbf} or becomes
non-commutative \cite{Berkooz:2022mfk}.
It is very interesting to understand the bulk dual of DSSYK
better. We leave these issues as interesting future problems.

\acknowledgments
This work was supported in part by 
JSPS Grant-in-Aid for Transformative Research Areas (A) 
``Extreme Universe'' 21H05187. KO was also supported 
by JSPS KAKENHI Grant 22K03594.

\appendix

\section{Alternative derivation of one-loop determinant}\label{app:alt}
In this appendix, we consider the diagonalization of the Hessian
$F_{ij}$ in the two-point function.
By the change of integration variables
$(\th_1,\th_2)\to(\th,x)$
\begin{equation}
\begin{aligned}
 \th_1&=\frac{\pi}{2}-\th+\hf\lap\tan x,\quad
\th_2&=\frac{\pi}{2}-\th-\hf\lap\tan x,
\end{aligned} 
\end{equation}
the two-point function is written as
\begin{equation}
\begin{aligned}
 \til{G}_2\sim \int \frac{d\th dx}{\cos^2x} e^{-\frac{1}{\la}F+h+\cO(\la)}
\end{aligned} 
\end{equation}
where
\begin{equation}
\begin{aligned}
 F&=2\th^2-\frac{4u}{\cos u}\sin\th+2\lap\left(\log\frac{\cos x}{\cos \th}
+x\tan x-\phi\frac{\cos\th}{\cos u}\tan x\right)\\
&\quad+\frac{\lap^2}{2}\left(1+\frac{u\sin\th}{\cos u}\right)\tan^2x
+\cO(\lap^3).
\end{aligned} 
\end{equation}
The saddle point solution is given by
\begin{equation}
\begin{aligned}
 \th_*&=u+\lap a+\cO(\lap^2),\quad x_*=\phi+\lap b-\lap a\phi\tan u+\cO(\lap^2),
\end{aligned} 
\end{equation}
with
\begin{equation}
\begin{aligned}
 a=-\frac{1+\phi\tan\phi}{2(1+u\tan u)},\quad
b=-\hf(1+u\tan u)\tan\phi.
\end{aligned} 
\end{equation}
We expand the integral around the saddle point as
\begin{equation}
\begin{aligned}
 \th=\th_*+\rt{\la}\vep,\quad x=x_*+\rt{\la}\Bigl(\lap^{-\hf}s-\vep\phi\tan u\Bigr),
\end{aligned} 
\end{equation}
where $\vep$ and $s$ parameterize the fluctuation around the saddle point.
Then, up to the quadratic order in the fluctuations $\vep,s$, we find
\begin{equation}
\begin{aligned}
 \til{G}_2\sim 
\frac{e^{-\frac{1}{\la}F_*+h_*}}{\cos^2x_*}\int d\vep ds e^{-F_2}
\end{aligned} 
\end{equation}
where
\begin{equation}
\begin{aligned}
 F_2&=\sec^2\phi s^2+2(1+u\tan u)\vep^2+\lap (A\vep^2+Bs^2)+\cO(\lap^2),
\end{aligned} 
\label{eq:F2-alt}
\end{equation}
with
\begin{equation}
\begin{aligned}
A&=2 a u-\phi^2 \tan ^2\phi \tan ^2u-\left(\phi^2-1\right) \tan ^2u+\phi \tan \phi+1,\\
B&=\frac{1}{2} \sec ^2\phi \left(\tan \phi (8 b-4 a \phi \tan u)+3 \tan ^2\phi (u \tan u+1)+u \tan
   u+1\right). 
\end{aligned} 
\end{equation}
As we can see from \eqref{eq:F2-alt}, the fluctuations $\vep$
and $s$ have no mixing at the quadratic order. 
Finally, at the one-loop level  we find
\begin{equation}
\begin{aligned}
 \til{G}_2\sim 
\frac{e^{-\frac{1}{\la}F_*+h_*}}{\cos^2x_*}
\frac{1-\lap A\bra\vep^2\ket-\lap B\bra s^2\ket}{\rt{\sec^2\phi (1+u\tan u)}},
\end{aligned}
\label{eq:alt-1loop} 
\end{equation}
where $\bra\vep^2\ket$ and $\bra s^2\ket$
are determined by the quadratic action \eqref{eq:F2-alt}
\begin{equation}
\begin{aligned}
 \bra\vep^2\ket=\frac{1}{4(1+u\tan u)},\quad
\bra s^2\ket=\hf\cos^2 \phi.
\end{aligned} 
\end{equation}
One can check that \eqref{eq:alt-1loop}
correctly reproduces the one-loop correction $\cA$ in \eqref{eq:A-2pt}.

\section{Summation formula}\label{app:sum}
We find the summation formula ($|\th|<\pi v$)
\begin{equation}
\begin{aligned}
 \sum_{|n|\geq1}\frac{(-1)^n\cos(\til{n}\th)}{\til{n}^2(\til{n}^2-1)}
&=-\frac{\th^2}{2}+\frac{2u^2}{3}+1-\frac{2u}{\sin 2u}\cos\th,\\
\sum_{|n|\geq1}\frac{(-1)^n\sin(\til{n}\th)}{\til{n}(\til{n}^2-1)}
&=\th-\frac{2u}{\sin2u}\sin\th,\\
\sum_{|n|\geq1}\frac{(-1)^n\cos(\til{n}\th)}{\til{n}^2-1}
&=1-\frac{2u}{\sin2u}\cos\th,\\
\sum_{|n|\geq1}\frac{(-1)^n\til{n}\sin(\til{n}\th)}{\til{n}^2-1}
&=-\frac{2u}{\sin2u}\sin\th,\\
\sum_{|n|\geq1}\frac{(-1)^n\til{n}^2\cos(\til{n}\th)}{\til{n}^2-1}
&=-\frac{2u}{\sin2u}\cos\th,
\end{aligned}
\label{eq:sum-nt} 
\end{equation}
where $\til{n}=n/v$.

For the out of time ordered correlator we discussed in section~\ref{sec:OTOC}, we also need
	\begin{align}
		\sum_{m=1}^\infty \frac{\cos(2m \th)}{4m^2 - v^2} \, = \, 
		\begin{dcases}
			\frac{1}{2v^2} \, - \, \frac{\pi^2 \cos(v \th - u)}{8u \sin u} \, , \qquad \qquad \ \, (0 < \th < \pi) \\[4pt]
			\frac{1}{2v^2} \, - \, \frac{\pi^2 \cos(v \th - 3u)}{8u \sin u} \, , \qquad \qquad (\pi < \th < 2\pi) 
		\end{dcases}
	\label{formu1}
	\end{align}
	\begin{align}
		\sum_{m=1}^\infty \frac{\cos(2m \th)}{m^2(4m^2 - v^2)} \, = \, 
		\begin{dcases}
			\frac{1}{v^2} \left( \frac{2}{v^2} \, - \, \frac{\pi^2 \cos(v \th - u)}{2u \sin u} \, - \, \th^2 + \pi \th - \frac{\pi^2}{6} \right) \, , \qquad \qquad (0 < \th < \pi) \\[4pt]
			\frac{1}{v^2} \left( \frac{2}{v^2} \, - \, \frac{\pi^2 \cos(v \th - 3u)}{2u \sin u} \, - \, \th^2 + 3\pi \th - \frac{13}{6} \pi^2 \right) \, , \qquad (\pi < \th < 2\pi) 
		\end{dcases}
	\end{align}
	\begin{align}
		\sum_{m=1}^\infty \frac{\cos((2m-1) \th)}{(2m-1)^2 - v^2} \, = \, 
		\begin{dcases}
		    \, - \, \frac{\pi^2 \sin(v \th - u)}{8u \cos u} \, , \qquad \qquad (0 < \th < \pi) \\[4pt]
			\frac{\pi^2 \sin(v \th - 3u)}{8u \cos u} \, , \qquad \qquad \ (\pi < \th < 2\pi) 
		\end{dcases}
	\label{formu3}
	\end{align}

\section{Zero-temperature factorization}
\label{app:factorization}
In Schwarzian theory, it is known that in zero temperature limit, the uncrossed four-point function factorizes into a product of two-point functions \cite{Mertens:2017mtv}.
In this subsection, we will show that this zero-temperature factorization is also true for the DSSYK for any value of $q$ (or $\lambda$).

For this purpose, we first define finite temperature correlators by
	\begin{align}
		G^\beta_2(\tau) \, &:= \, \til{G}_2(\tau, \beta-\tau) \, , \\
		G^\beta_4(\tau_1, \tau_2, \tau_3, \tau_4) \, &:= \, \til{G}_4(\tau_{12}, \tau_{23}, \tau_{34}, \tau_{41}+\beta) \, ,
	\end{align}
where the RHS' are defined in (\ref{eq:G2_intro}) and (\ref{eq:G4_intro}). Each $\tau_i$ represents the matter operator insertion time.
We also shift the energy as
	\begin{align}
		E(\th) \, \Rightarrow \, \frac{2}{\sqrt{1-q}} \, (1- \cos\th) \, .
	\end{align}
This sets the ground state energy $E(0)=0$. 

Let us first study the zero-temperature limit of the two-point function:
	\begin{align}
		G^\beta_2(\tau) \, = \, \int_0^\pi \frac{d\theta_1}{2\pi} \frac{d\theta_2}{2\pi} \, \mu(\theta_1) \mu(\theta_2) e^{-\tau E(\theta_1)} e^{(\tau-\beta) E(\theta_2)} \, 
        \frac{(e^{-2\Delta}; q)_\infty}{(e^{-\Delta+i(\pm \theta_1 \pm \theta_2)}; q)_\infty} \, .
	\end{align}
Due to the Boltzmann factor $e^{-\beta E(\theta_2)}$, the contribution to the $\theta_2$ integral is localized to the ground state, i.e. $\theta_2 \to 0$.
In this limit, we have
	\begin{align}
		\mu(\th) \, = \, -4 (q;q)_\infty^3 \, \sin^2\theta  \, + \, \cdots \, .
	\end{align}
Therefore, the zero-temperature two-point function is given by
	\begin{align}
		G^\infty_2(\tau) \, = \, - \frac{2\sqrt{1-q}}{\beta} \, (q;q)_\infty^3 \, e^{-\frac{2\beta}{\sqrt{1-q}}} I_1\left( \frac{2\beta}{\sqrt{1-q}} \right) 
		\int_0^\pi \frac{d\theta}{2\pi} \, \mu(\theta) e^{-\tau E(\theta)} \, \frac{(e^{-2\Delta}; q)_\infty}{(e^{-\Delta \pm i\theta}; q)_\infty^2} \, .
	\end{align}
Before studying the four-point function, let us here consider the late time behavior of this zero-temperature two-point function.
Evaluating the late time behavior by the same method as zero-temperature limit discussed above, we find
	\begin{align}
		\lim_{\tau \to \inf} G^\infty_2(\tau) \, = \, \frac{4(1-q)}{\beta \tau} \, e^{-\frac{2(\beta+\tau)}{\sqrt{1-q}}} \,
        (q;q)_\infty^6 \frac{(e^{-2\Delta}; q)_\infty}{(e^{-\Delta}; q)_\infty^4} \, I_1\left( \frac{2\beta}{\sqrt{1-q}} \right) I_1\left( \frac{2\tau}{\sqrt{1-q}} \right) \, .
	\end{align}
Since 	
	\begin{align}
		\lim_{\tau \to \infty} I_1\left( \frac{2\tau}{\sqrt{1-q}} \right) \, = \, \frac{(1-q)^{\frac{1}{4}}}{\sqrt{4\pi \tau}} \, e^{\frac{2\tau}{\sqrt{1-q}}} \, + \, \cdots \, ,
	\end{align}
The late time behavior of the zero-temperature two-point function is $G^\infty_{\Delta}(\tau) \propto \tau^{-3/2}$, which agrees with the late time behavior in Schwarzian theory.

Now we study zero-temperature limit of the four-point function:
	\begin{align}
		G^\beta_4 \, &= \, \int_0^\pi \prod_{i=1,3} \left( \frac{d\theta_i}{2\pi}  \, \mu(\theta_i) e^{-(\tau_i - \tau_{i+1}) E(\theta_i)} (e^{-2\Delta_i}; q)_\infty \right) \nn\\
		&\qquad \times \int_0^\pi \frac{d\theta_2}{2\pi} \mu(\th_2) e^{-(\tau_{23} + \tau_{41}+\beta) E(\theta_2)} \prod_{j=1}^2 \frac{1}{(e^{-\Delta+i(\pm \theta_j \pm \theta_{j+1})}; q)_\infty} \, ,
	\end{align}
where $\Delta_3=\Delta_2$.
Again, zero-temperature limit $\beta\to \infty$ localizes $\theta_2 \to 0$. Therefore, we find
	\begin{align}
		G^\beta_4 \, = \, - \frac{\beta}{2\sqrt{1-q}} \, (q;q)_\infty^{-3} \, e^{\frac{2\beta}{\sqrt{1-q}}} \left( I_1\left( \frac{2\beta}{\sqrt{1-q}} \right) \right)^{-1} G^\infty_2(\tau_{12}) \, G^\infty_2(\tau_{34}) \, .
	\end{align}
We note that as in Schwarzian theory, the zero-temperature four-point function can be factorized only in this $s$-channel, but not in the $t$ or $u$-channels, which is obvious from the matter operator contractions.

\bibliography{refs}
\bibliographystyle{utphys}

\end{document}